\documentclass[pre,twocolumn,amsmath,amssymb,superscriptaddress]{revtex4-1}

\usepackage{minitoc}

\usepackage{mathtools,amscd}

\usepackage{xr}
\usepackage{url}
\usepackage{hyperref}
\hypersetup{
    colorlinks,
    citecolor=blue,
    filecolor=blue,
    linkcolor=blue,
    urlcolor=blue
}

\usepackage{tabularx}

\usepackage{mathrsfs,ragged2e}
\usepackage{verbatim}
\usepackage{graphicx}	
\usepackage{dcolumn}	
\usepackage{bm}		
\usepackage{bbm,upgreek}
\usepackage{mathtools}
\usepackage{physics}

\usepackage{xr}
\usepackage{color}
\usepackage[cal=boondoxo]{mathalfa}
\usepackage{palatino}

\makeatletter
\def\maketitle{
\@author@finish
\title@column\titleblock@produce
\suppressfloats[t]}
\makeatother

\usepackage{graphicx}

\usepackage{dsfont}
\usepackage{cancel}
\usepackage{xcolor}
\usepackage{harpoon}
\usepackage{accents}

\newcommand*{\Scale}[2][4]{\scalebox{#1}{$#2$}}%
\newcommand*{\inliner}[1]{{\smash{\Scale[0.9]{#1}}}}

\newcommand*{\idea}[1]{\medskip \noindent \textbf{#1}}

\begin{document}

\title{The one-message-per-cell-cycle rule: \\ A conserved minimum transcription level for essential genes}


\author{Teresa W. Lo}
\author{Han Kyou James Choi}
\author{Dean Huang}
\affiliation{Department of Physics, University of Washington, Seattle, Washington 98195, USA}
\author{Paul A. Wiggins}
\email{pwiggins@uw.edu}
\affiliation{Department of Physics, University of Washington, Seattle, Washington 98195, USA}
\affiliation{Department of Bioengineering, University of Washington, Seattle, Washington 98195, USA}
\affiliation{Department of Microbiology, University of Washington, Seattle, Washington 98195, USA}

\begin{abstract}


The inherent stochasticity of cellular processes leads to significant cell-to-cell variation in protein abundance. Although this noise has already been characterized and modeled, its broader implications and significance remain unclear. In this paper, we revisit the noise model and identify the number of messages transcribed per cell cycle as the critical determinant of noise. In yeast, we demonstrate that this quantity predicts the non-canonical scaling of noise with protein abundance, as well as quantitatively predicting its magnitude. We then hypothesize that growth robustness requires an upper ceiling on noise for the expression of essential genes, corresponding to a lower floor on the transcription level. We show that just such a floor exists: a minimum transcription level of one message per cell cycle is conserved between three model organisms: \textit{Escherichia coli}, yeast, and human. Furthermore, all three organisms transcribe the same number of messages per gene, per cell cycle. This common transcriptional program reveals that robustness to noise plays a central role in determining the expression level of a large fraction of essential genes, and that this fundamental optimal strategy is conserved from \textit{E.~coli} to human cells.

\end{abstract}



\maketitle


\section*{Introduction}

All molecular processes are inherently stochastic on a cellular scale, including the processes of the central dogma, responsible for gene expression \cite{Raser:2005we,phillips2013physical}.  As a result, the expression of every protein is subject to cell-to-cell variation in abundance \cite{Raser:2005we}.
 Many interesting proposals have been made to describe the potential biological significance of this noise, including bet-hedging strategies, the necessity of feedback in gene regulatory networks, \textit{etc} \cite{Raser:2005we,Paulsson:2000xi,Paulsson:2004zy}. However, it is less clear to what extent noise plays a central role in determining the function of the gene expression process more generally. For instance, Hausser \textit{et al.} have described how the tradeoff between economy (\textit{e.g.}~minimizing the number of transcripts) and precision (minimizing the noise) explains why genes with high transcription rates and low translation rates are not observed \cite{Hausser:2019fi}. Although these results suggest that noise may provide some coarse limits on the function of gene expression, this previous work does not directly address a central challenge posed by noise: How does the cell ensure that the lowest expression essential genes, which are subject to the greatest noise, have sufficient abundance in all cells for robust growth?

To investigate this question, we first 
focus on noise in \textit{Saccharomyces cerevisiae} (yeast), and find that the noise scaling with protein abundance is not canonical. We re-analyze the canonical stochastic kinetic model for gene expression, the telegraph model \cite{Raj:2006ww,Iyer-Biswas:2009sd,Peccoud1995}, to understand the relationship between the underlying kinetic parameters and the distribution of protein abundance in the cell.
As previously reported, we find that the protein abundance for a gene is described by a gamma distribution with two parameters: the \textit{message number}, defined as the total gene message number transcribed per cell cycle, and the translation efficiency, which is the mean protein number translated per message. 
Protein expression noise is completely determined by the message number \cite{Paulsson:2000xi,Friedman:2006oh}. 
Although these results have been previously reported, the distinction between message number \textit{per cell} versus \textit{per cell cycle} and even between \textit{mean protein number} and \textit{mean message number} is often neglected (\textit{e.g.}~\cite{Bar-Even:2006rv}). 

To explore the distinction between these parameters and provide clear evidence of the importance of the message number, we return to the analysis of noise in yeast.
In yeast, the translation efficiency increases with message number \cite{Weinberg:2016ll}.
By fitting an empirical model for the translation efficiency, we demonstrate that the noise should scale with a half-power of protein abundance. We demonstrate that this non-canonical scaling is observed and that our translation model makes a parameter-free prediction for the noise. The prediction is in close quantitative agreement with observation \cite{Newman:2006nl}, confirming that the message number is the key determinant of noise strength.

\begin{figure}
  \centering
   \includegraphics[width=0.40\textwidth]{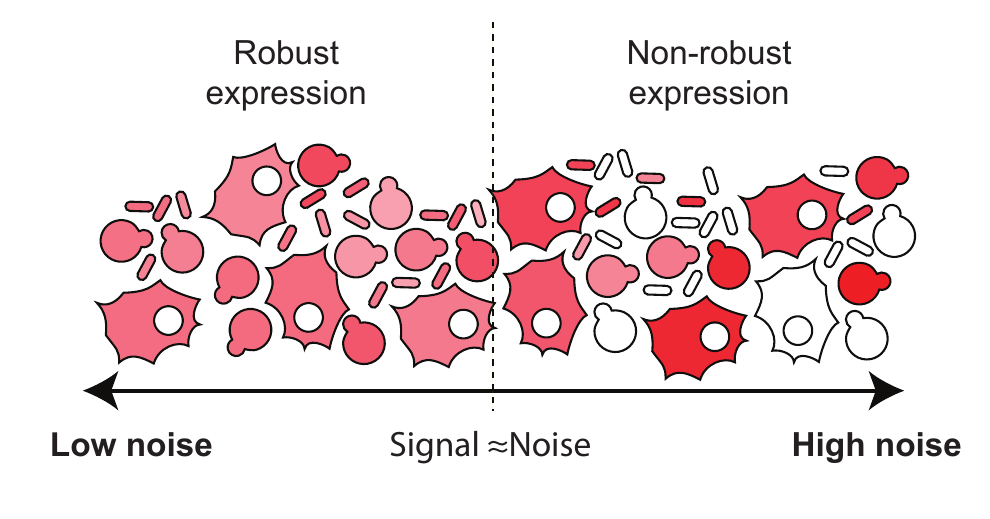}
      \caption{\textbf{Robustness hypothesis:} The stochasticity in gene expression is represented by the red shading. We hypothesize that robust growth requires sufficiently low noise levels for cellular function. We hypothesize that this critical noise level should be below the level where the signal (mean) equals noise (standard deviation).}
\end{figure}


Finally, we use this result to explore the hypothesis that there is a minimum expression level for essential genes, dictated by noise.  The same mean expression level can be achieved by a wide range of different translation and transcription rates with different noise levels. We hypothesize that growth robustness requires that essential genes (but not non-essential genes) are subject to a floor expression level, below which there is too much cell-to-cell variation to ensure  growth. 
To test this prediction, we analyze transcription in three model organisms, \textit{Escherichia coli}, yeast, and \textit{Homo sapiens} (human), with respect to three related gene characteristics: transcription rate, cellular message number, and message number per cell cycle. 
As predicted by the noise-based mechanism, we observe an organism-independent floor for the number of messages transcribed per cell cycle for essential genes, but not non-essential genes. We conclude that virtually all essential genes are transcribed at a rate of at least once per cell cycle. This analysis strongly supports the hypothesis 
that the same biological optimization imperatives, which determine the transcription rates of many low-expression genes, are conserved from \textit{E.~coli} to human.




\section*{Results}

\idea{Implications of noise on growth robustness.} 
With the realization of the stochasticity of central dogma processes, a key question is how cells can grow robustly in spite of cell-to-cell variations in protein expression. The noise in protein abundance is defined as the coefficient of variation squared \cite{Elowitz:2002tb,Swain:2002te,Newman:2006nl}:
\begin{equation}
{\rm CV}^2_p \equiv \textstyle\frac{\sigma^2_p}{\mu^2_p},
\end{equation}
where $\sigma^2_p$ is the variance of protein number and $\mu_p \equiv \overline{N}_p$ its mean. It is important to emphasize that protein abundance must double between birth and cell division in symmetrically dividing cells during steady state growth. The protein abundance should therefore be interpreted either as expression per unit volume \cite{Taniguchi2010} or the abundance associated with cells of a defined volume \cite{Newman:2006nl}.

The coefficient of variation is inversely related to protein abundance and therefore low-copy proteins have the highest noise \cite{Elowitz:2002tb,Swain:2002te,Paulsson:2000xi,Friedman:2006oh,Taniguchi2010,Newman:2006nl}. The challenge faced by the cell is that many essential proteins, strictly required for cell growth, are relatively low abundance. How does the cell ensure sufficient protein abundance in spite of cell-to-cell variation in protein number?
It would seem that growth robustness demands that, for essential proteins, the mean should be greater than the standard deviation: 
\begin{equation}
{\rm CV}^2_p < 1, \label{eq:noiseceil}
\end{equation}
 in order to ensure that protein abundance is sufficiently high enough to avoid growth arrest. To what extent do essential proteins obey this noise threshold?

\idea{What determines the strength of the noise?} 
Usually, noise is argued to be proportional to inverse protein abundance (\textit{e.g.}~\cite{Paulsson:2000xi,Paulsson:2004zy,Bar-Even:2006rv}):
\begin{equation}
{\rm CV}^2_p \propto \mu_p^{-1}, \label{eqn:nullmodel}
\end{equation}
for low abundance proteins, motivated both by theoretical and experimental results \cite{Bar-Even:2006rv,Taniguchi2010} and in some cases obeying a low-translation efficiency limit \cite{Taniguchi2010}:
\begin{equation}
{\rm CV}^2_p \approx \mu_p^{-1}. \label{eqn:nullmodelapprox2}
\end{equation}
Can this model be used to make quantitative predictions of the noise? \textit{E.g.}, is the scaling of Eq.~\ref{eqn:nullmodel} correct? Can the coefficient of proportionality be predicted?
Although Eq.~\ref{eqn:nullmodel} appears to describe \textit{E.~coli} quite well \cite{Taniguchi2010}, the situation in yeast is more complicated \footnote{Although there have been claims that Eq.~\ref{eqn:nullmodel} is consistent with the data \cite{Bar-Even:2006rv}, these authors did not fit competing models, nor did they perform a proteome-wide analysis of protein abundance and noise.}. To analyze the statistical significance of the deviation from the canonical noise model in yeast, we can fit an empirical model to the noise \cite{Elowitz:2002tb,Swain:2002te}:
\begin{equation}\label{eqn:noiseempirical}
{\rm CV}_p^2 = \textstyle\frac{b}{\mu_p^a}+c. 
\end{equation}
In the null hypothesis, $a=1$ (canonical scaling), while $b$ and $c$ are unknown parameters. $c$ corresponds to the noise floor. In the alternative hypothesis, all three coefficients are unknown. (A detailed description of the statistical model is given in the Supplemental Material Sec.~\ref{sec:SMnoise}.)

\begin{figure}
  \centering
   \includegraphics[width=0.48\textwidth]{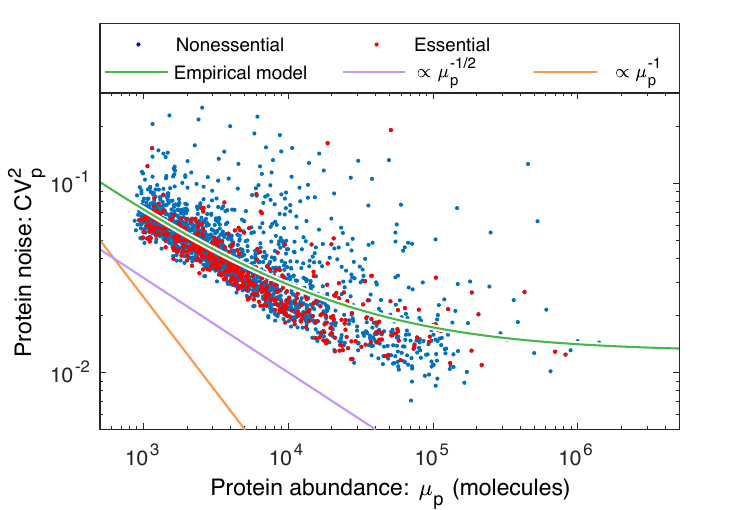}
      \caption{\textbf{A non-canonical scaling is observed for gene-expression noise in yeast.} The protein expression noise (${\rm CV}^2_p$) for yeast scales like $\inliner{\mu_p^{-1/2}}$ (purple) rather than the canonical $\inliner{\mu_p^{-1}}$ (orange) for low-abundance proteins. (Data from Ref.~\cite{Newman:2006nl}.) An empirical noise model (Eq.~\ref{eqn:noiseempirical}, green) fit to the essential genes gives an estimate of the protein-abundance scaling of $\mu_p^{-0.57}$. 
      \label{fig:empmodnoiseyeast}}
\end{figure}

The canonical model fails to fit the noise data for yeast as reported by Newman \textit{et al.} \cite{Newman:2006nl}: The null hypothesis is rejected with p-value $p = 6\times 10^{-36}$. The model fit to the data is shown in Fig.~\ref{fig:empmodnoiseyeast}. The estimated scaling exponent for protein abundance in the alternative hypothesis is $a = 0.57\pm 0.02$, and a detailed description of the statistical model and parameter fits is provided in Supplementary Material Sec.~\ref{sec:yeastnoiseempiricalmodel}. As shown in Fig.~\ref{fig:empmodnoiseyeast}, even from a qualitative perspective, the scaling of the yeast noise at low copy number is much closer to $\inliner{\mu_p^{-1/2}}$ than to canonical assumption $\mu_p^{-1}$ (Eq.~\ref{eqn:nullmodel}). In particular, above the detection threshold, the noise is always larger than the low-translation efficiency limit (Eq.~\ref{eqn:nullmodelapprox2}).

\begin{figure*}
  \centering
   \includegraphics[width=0.95\textwidth]{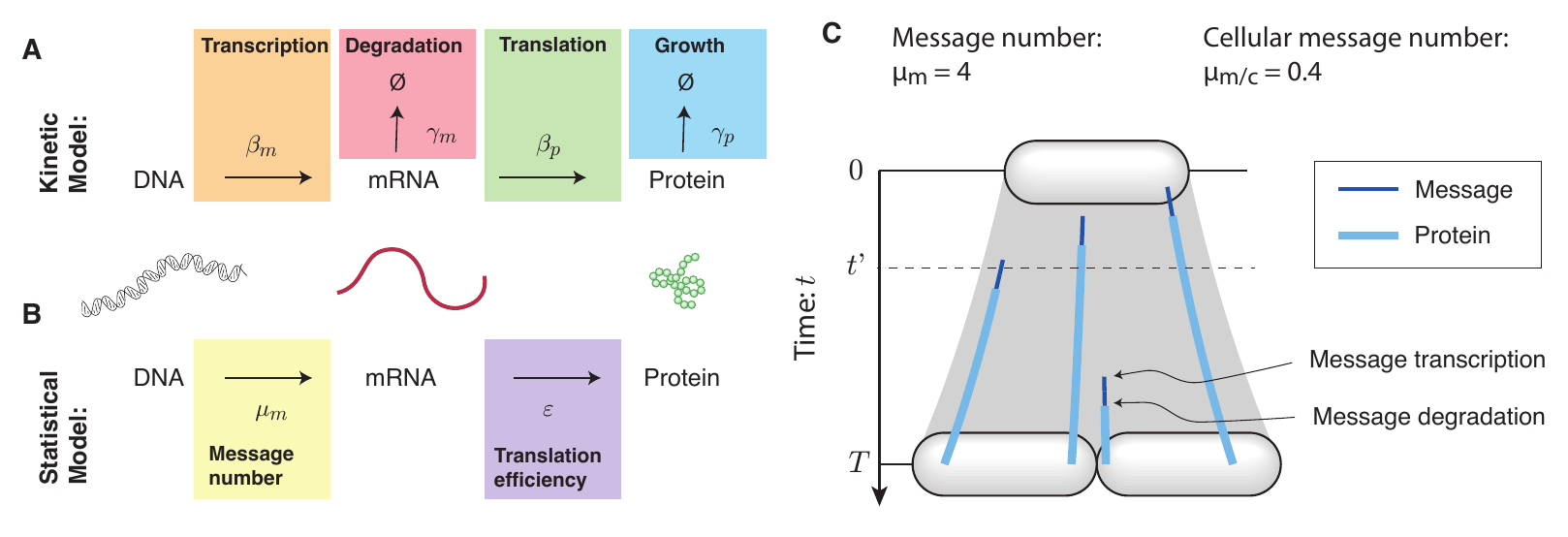}
      \caption{\textbf{Panel A: Kinetic model for the central dogma.} The telegraph model is a stochastic kinetic model for protein synthesis, described by four gene-specific rate constants: the transcription rate ($\beta_m$), the message degradation rate ($\gamma_m$), the translation rate ($\beta_p$), and the dilution rate ($\gamma_p$).  \textbf{Panel B: Statistical model for the central dogma.} The predicted distribution in protein abundance is described by a gamma distribution, which is parameterized by two unitless constants: the shape parameter $\mu_m$, the mean number of messages transcribed per cell cycle, and the scale parameter $\varepsilon$, the mean number of proteins translated per message.  \textbf{Panel C: Message number.} The  \textit{message number} ($\mu_m$) is defined as the mean total number of messages (dark blue) transcribed per cell cycle. Here, four total messages  are transcribed and translated to protein (light blue); however, due to message degradation, at time $t'$, only one message is present in the cell.  Cellular message number ($\mu_{m/c}$) is defined as the mean number of messages per cell at time $t$. 
      \label{fig:schem}}
\end{figure*}


\idea{Stochastic kinetic model for central dogma.} To understand the failure of the canonical assumptions, we revisit the underlying model. The telegraph model for the central dogma describes multiple steps in the gene expression process:
 Transcription generates mRNA messages \cite{Crick:1970vy}. These messages are then translated to synthesize the protein gene products \cite{Crick:1970vy}.  Both mRNA and protein are subject to degradation and dilution \cite{Hargrove:1989vo}. (See Fig.~\ref{fig:schem}A.) At the single cell level, each of these processes are stochastic. We will model these processes with the stochastic kinetic scheme   \cite{Crick:1970vy}:
\begin{equation}
  \begin{CD}
   {\rm DNA} @>\beta_m>> {\rm mRNA} & @>\beta_p>>  {\rm Protein} \\
     & & @V\gamma_mVV & @V\gamma_pVV\\
     & & \O && &  \; \O, \label{stochmodel}
  \end{CD}
\end{equation}
where $\beta_m$ is the transcription rate (s$^{-1}$), $\beta_p$ is the translation rate (s$^{-1}$), $\gamma_m$ is the message  degradation rate (s$^{-1}$), and $\gamma_p$ is the  protein effective degradation rate (s$^{-1}$).  The message lifetime is $\tau_m\equiv \gamma_m^{-1}$. For most protein in the context of rapid growth, dilution is the dominant mechanism of protein depletion and therefore $\gamma_p$ is approximately the growth rate  \cite{KOCH:1955oa, Martin-Perez:2017jx, Taniguchi2010}: $\gamma_p = T^{-1}\ln 2$,  where $T$ is the doubling time. We will discuss a more general scenario below.


\idea{Statistical model for  protein abundance.} To study the stochastic dynamics of gene expression, we used a stochastic Gillespie simulation \cite{Gillespie1992,Gillespie1977}. (See Supplemental Material Sec.~\ref{SecGS}.) In particular, we were interested in the explicit relation between the kinetic parameters $(\beta_m, \gamma_m, \beta_p, \gamma_p)$ and experimental observables.

Consistent with previous reports \cite{Paulsson:2000xi,Friedman:2006oh}, we find that the distribution of protein number per cell (at cell birth) was described by a gamma distribution:
$N_p \sim \Gamma(\theta_{\Gamma},k_{\Gamma})$, where $N_p$ is the protein number at cell birth and $\Gamma$ is the gamma distribution which is parameterized by a scale parameter $\theta_{\Gamma}$ and a shape parameter $k_{\Gamma}$. 
(See Supplementary Material Sec.~\ref{SecGS}.)
The relation between the four kinetic parameters and these two statistical parameters has already been reported, and have clear biological interpretations \cite{Friedman:2006oh}: The scale parameter: 
\begin{eqnarray} 
\theta_{\Gamma} = \varepsilon \ln 2, \label{eqn:eff_rates} 
\end{eqnarray}
is proportional to the translation efficiency:
\begin{eqnarray}
\varepsilon \equiv \textstyle \frac{\beta_p}{\gamma_m}, \label{eqn:TE}
\end{eqnarray} 
where  $\beta_p$ is the translation rate and $\gamma_m$ is the message degradation rate.   $\varepsilon$ is understood as the mean number of proteins translated from each message transcribed. 
The shape parameter $k_{\Gamma}$ can also be expressed in terms of the kinetic parameters \cite{Friedman:2006oh}: 
\begin{equation}
k_{\Gamma} = \textstyle\frac{\beta_m}{\gamma_p }; \label{eqn:shape2}
\end{equation}
however, we will find it more convenient to express the scale parameter in terms of the cell-cycle message number:
\begin{eqnarray}
    \mu_{m} \equiv \beta_m T =  k_\Gamma \ln 2,
\end{eqnarray}
which can be interpreted as the mean  number of messages transcribed per cell cycle. Forthwith, we will abbreviate this quantity \textit{message number} in the interest of brevity.


\begin{table*}
\resizebox{\textwidth}{!}{\begin{tabularx}{1. \textwidth}{ X | X | X  X X | X  X  X | X X }
                         &            &                   &                             &                               &  \multicolumn{3}{c|}{Total number of }       &    \multicolumn{2}{c}{Average}      \cr
\centering{Model organism}                &  \centering{Growth condition} & \centering{Doubling time}:  & \centering{Message lifetime}: &  \centering{Message recycling ratio:} & \centering{messages /cell:}      &   \centering{messages /cell-cycle:}          &    \centering{proteins:}     &  \centering{translation efficiency:}  &  \centering{translation rate:} \cr
                         &                               & \centering{$T$}    &  \centering{$\tau_m$} & \centering{ $T/\tau_m$ }             &   \centering{$N^{\rm tot}_{m/c}$}  &   \centering{$N^{\rm tot}_{m}$}    &  \centering{$N^{\rm tot}_{p}$}   & \centering{$\varepsilon$}  &  \centering{$\beta_p$ (h$^{-1}$)}  \cr
\hline
\hline
 \textit{Escherichia coli}  & \raggedleft   LB & \raggedleft  $30$ min   & \raggedleft $2.5$ min  & \raggedleft $12$  &  \raggedleft $7.8 \times 10^3$    & \raggedleft $9.4 \times 10^4$ & \raggedleft $3 \times 10^6$  & \raggedleft  $32$  & \raggedleft  $770$   \cr 
 (\textit{E.~coli})         & \raggedleft   M9 & \raggedleft  $90$ min  & \raggedleft $2.5$ min   & $36$ \raggedleft & \raggedleft $2.4 \times 10^3$  & \raggedleft $8.6 \times 10^4$ & \raggedleft $3 \times 10^6$   & \raggedleft $35$ & \raggedleft  $833$  \cr 
                     \hline
 \textit{Sacchromyces cerevisiae (Yeast--haploid)}     & \raggedleft  YEPD    & \raggedleft  $90$ min  & \raggedleft  $22$ min  & \raggedleft $4$ & \raggedleft $2.9 \times 10^4$  & \raggedleft $1.2 \times 10^5$ & \raggedleft  $5 \times 10^7$  & \raggedleft $420$   & \raggedleft $1100$  \cr 
                                          \hline
 \textit{Homo sapiens (Human)}         & \raggedleft   Tissue   & \raggedleft  $24$ h    & \raggedleft  $14$ h   & \raggedleft  $1.7$ & \raggedleft   $3.6 \times 10^5$  & \raggedleft $6.2 \times 10^5 $  & \raggedleft $2 \times 10^9$  & \raggedleft  $3.2\times 10^3$ & \raggedleft  $230$  \cr 
                     \hline
\end{tabularx}}

\caption{ \textbf{Central dogma parameters for three model organisms.}  Columns three through seven hold representative values for  measured central-dogma parameters for the model organisms described in the paper. \label{tab1} The sources of the numbers and estimates are described in the Supplemental Material Sec.~\ref{sec:desriptonTab1}.}
\end{table*}

In terms of two gamma parameters, the mean and the squared coefficient of variation are:
\begin{eqnarray}
         \label{eqn:meanprotnum}
\mu_p  &=& k_\Gamma \theta_\Gamma  =  \textstyle \mu_{m} \varepsilon \,\label{eqn:protmean} \\ 
{\rm CV}^2_p &=&  \textstyle \frac{1}{k_\Gamma}  =  \textstyle \frac{\ln 2}{\mu_{m}}, \label{eqn:protnoise}
\end{eqnarray}
where the noise depends on the message number ($\mu_m$), not the mean protein number ($\mu_p$). (Eq.~\ref{eqn:protnoise} only applies when $\varepsilon \gg 0$ \cite{Paulsson:2000xi,Friedman:2006oh}.) 
Are these theoretical results consistent with the canonical model (Eq.~\ref{eqn:nullmodel})? 
We can rewrite the noise in terms of the protein abundance and translation efficiency:
\begin{equation}
 {\rm CV}_p^2 = \textstyle \frac{\varepsilon \ln 2}{\mu_p},  \label{eq:noiseTE}
\end{equation}
which implies that the canonical model only applies when the translation efficiency ($\varepsilon$) is independent of expression ($\mu_p$).

\idea{Measuring the message number.} The prediction for the noise (Eq.~\ref{eqn:protnoise}) depends on the message number ($\mu_m$). However, mRNA abundance is typically characterized by a closely related, but distinct quantity: Quantitative RNA-Seq and methods that visualize fluorescently-labeled mRNA molecules typically measure the  number of messages per cell \cite{Raj:2006ww}.  We will call the mean of this number the \textit{cellular message number} $\mu_{m/c}$. In the kinetic model, these different message abundances are related:  
\begin{equation}
\mu_{m} = \textstyle\frac{T}{\tau_m} \mu_{m/c}, \label{eqn:cellularmn}
\end{equation}
by the message recycling  ratio, $T/\tau_m$, which can be interpreted as the average number of times messages are recycled during the cell cycle. 
To estimate the message number, we will scale the observed cellular message number $\mu_{m/c}$ by the message recycling ratio, using the mean message lifetime. Fig.~\ref{fig:schem}C illustrates the difference between the message number and the cellular message number. The mean lifetimes, message recycling ratios, as well as the total message number for three model organisms are shown in Tab.~\ref{tab1}.

\begin{figure}
  \centering
   \includegraphics[width=0.48\textwidth]{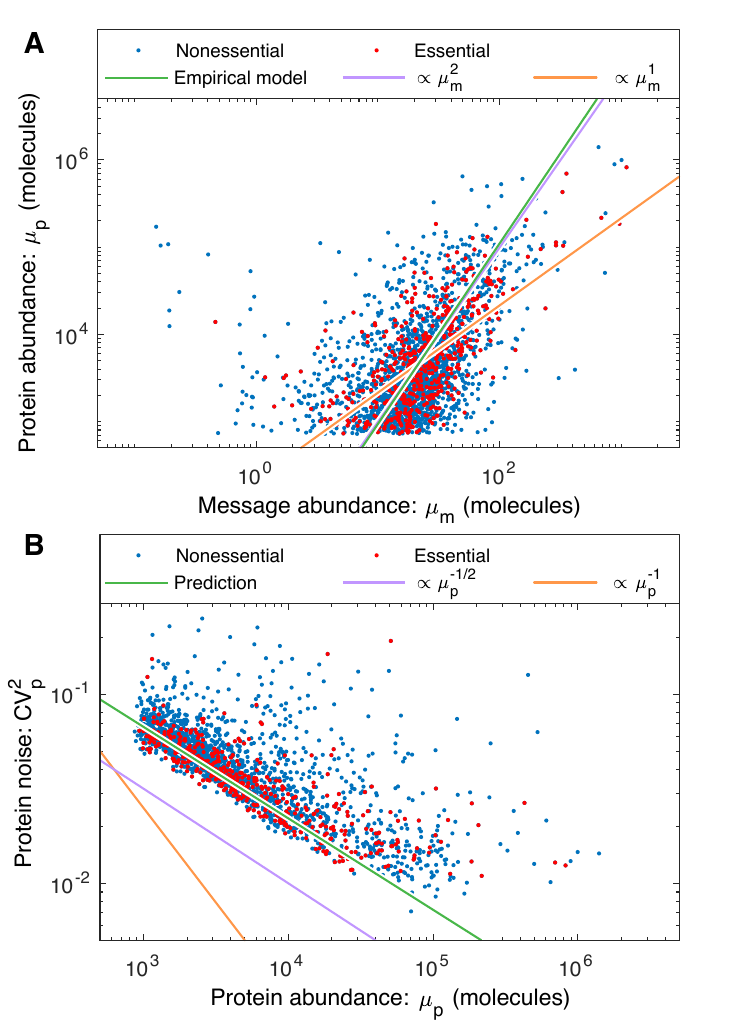}
      \caption{ \textbf{Panel A: An empirical model for protein number $\mu_p$ in yeast.} The canonical noise model assumes constant translation efficiency, which would imply that protein number is proportional to the message number (orange); however, the empirical fit (green) shows that protein number scales close to the square of message number (purple): $\inliner{\mu_p\propto \mu_m^2}$. The protein abundance has a cutoff near $10^1$ due to the autofluorescence cutoff \cite{Newman:2006nl}.
      \label{fig:transEff}
  \textbf{Panel B: The statistical noise model predicts the observed noise.} The statistical noise model (Eq.~\ref{eqn:protnoise}) and empirical model for protein number (Eq.~\ref{eqn:empmod}) make a parameter-free prediction of the noise (green). This prediction both closely matches the observed scaling ($\inliner{\propto \mu^{-1/2}_p}$, purple) relative to the canonical scaling ($\inliner{\propto \mu^{-1}_p}$, orange) and quantitatively estimates magnitude (vertical offset). This prediction does not include the contribution of noise floor, relevant for describing high-expression proteins. 
      \label{fig:noisefig} }
\end{figure}

\idea{Construction of an empirical model for protein number.} To model the noise as a function of protein abundance ($\mu_p$), we will determine the empirical relation between mean protein levels and message abundance by fitting to  Eq.~\ref{eqn:meanprotnum}. 
Note that the objective here is only to estimate $\mu_m$ from $\mu_p$, not to model the process mechanistically (\textit{e.g.}~\cite{Shah:2013ws}.)
The message numbers are estimated from RNA-Seq measurements, scaled as described above (Eq.~\ref{eqn:cellularmn}). The protein abundance numbers come from fluorescence and mass-spectrometry based assays \cite{Newman:2006nl,Godoy:2008eb}, with overall normalization chosen to match reported total cellular protein content. (See Supplemental Material Sec.~\ref{sec:MS}.) The resulting fit generates our empirical translation model for yeast:
\begin{eqnarray} 
\mu_p &=& 8.0\  \mu_m^{2.1}, \label{eqn:empmod}
\end{eqnarray}
where both means are in units of molecules. 
(An error analysis for both model parameters is described in Supplementary Material Sec.~\ref{sec:empmodel0}.) The data and model are shown in Fig.~\ref{fig:transEff}A.




\idea{Prediction of the noise scaling with abundance.} 
Now that we have fit an empirical model that relates $\mu_p$ and $\mu_m$, we return to the problem of predicting the yeast noise. We apply the relation (Eq. \ref{eqn:empmod}) to Eq.~\ref{eqn:protnoise} to make a parameter-free prediction of the noise as a function of protein abundance:
\begin{equation}
{\rm CV}_p^2 = 1.9\  \mu_p^{-0.48}. \label{eqn:noisemodelstat}
\end{equation}
An error analysis for both model parameters is described in Supplementary Material Sec.~\ref{sec:empmodel3}. Our noise model (Eq.~\ref{eqn:noisemodelstat})  makes both a qualitative and quantitative prediction: (i) From a qualitative perspective, the model suggests that the  $\mu_p$ exponent should be roughly $\textstyle \frac{1}{2}$ for yeast, rather than the canonically assumed scaling exponent of 1. (ii) From a quantitative perspective, the model also predicts the coefficient of proportionality if the empirical relation between protein and message abundances  is known (Eq.~\ref{eqn:empmod}). 

\idea{Observed noise in yeast matches the predictions of the empirical model.} Newman \textit{et al.} have characterized protein noise by flow cytometry of strains expressing fluorescent fusions expressed from their endogenous promoters \cite{Newman:2006nl}. The comparison of this data to the prediction of the statistical expression model (Eq.~\ref{eqn:noisemodelstat}) are shown in Fig.~\ref{fig:noisefig}. From a qualitative perspective, the predicted scaling exponent of $-0.48$ comes very close to capturing the scaling of the noise, as determined by the direct fitting of the empirical noise model (Eq.~\ref{eqn:noiseempirical} and Fig.~\ref{fig:empmodnoiseyeast}). From a quantitative perspective, the predicted coefficient of Eq.~\ref{eqn:noisemodelstat} also fits the observed noise.

From both the statistical analysis (Eq.~\ref{eqn:noiseempirical}) and visual inspection (Fig.~\ref{fig:noisefig}C), it is clear that the noise in yeast does not obey the canonical model (Eq.~\ref{eqn:nullmodel}). However, the noise in \textit{E.~coli} does obey the canonical model for low copy messages \cite{Taniguchi2010}. (See Fig.~\ref{fig:noisefig}C.) 
Why does the noise scale differently in the two organisms? The key difference is that the empirical relation between the protein and message numbers are different. In \textit{E.~coli}, $\mu_p \propto \mu_m^1$ \cite{Balakrishnan:2022ai}. Our analysis therefore predicts the canonical model (Eq.~\ref{eqn:nullmodel}) should hold for \textit{E.~coli}, but not for yeast, as illustrated schematically in Fig.~\ref{fig:transEff}.
(Additional discussion can be found in the Supplementary Material Sec.~\ref{sec:noiseEcoli}.).

\begin{figure}
  \centering
   \includegraphics[width=0.48\textwidth]{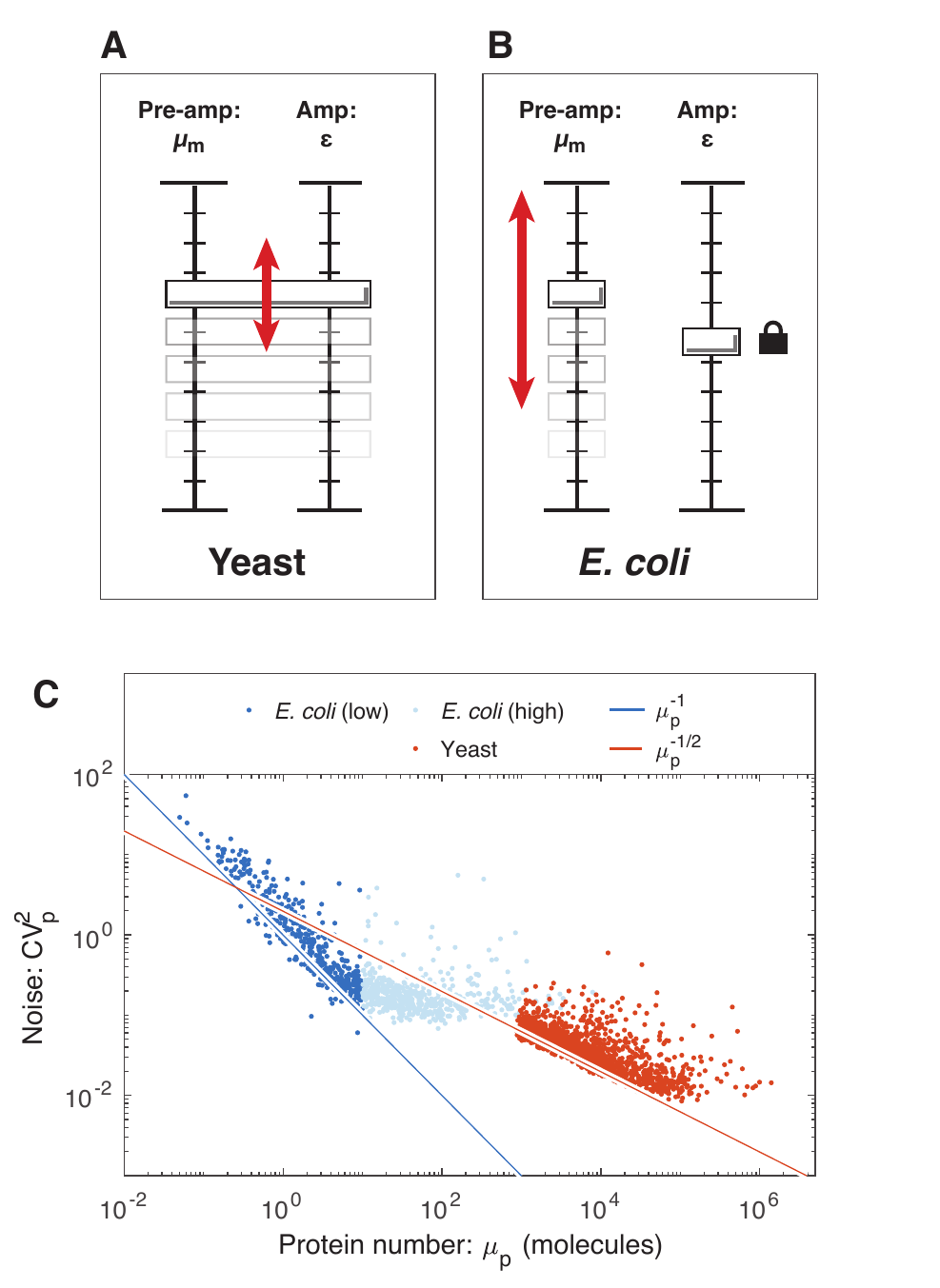}
      \caption{ \textbf{Understanding the distinct central dogma strategies using the amplifier analogy. Panel A: Yeast.} High expression ($\mu_p$) is typically achieved by coordinated small increases in both transcription ($\mu_m$) and translation ($\varepsilon$), relative to low-expression genes.
     \textbf{Panel B: \textit{E.~coli}.}  High expression ($\mu_p$) is typically achieved by a large increase in  transcription ($\mu_m$) only, relative to low-expression genes. Translation ($\varepsilon$) is uncorrelated.
     \textbf{Panel C: Distinct noise scaling with gene expression.} Due to the coordinated changes in both transcription and translation in yeast, noise scaling is weaker than in \textit{E.~coli}, where only transcription changes. The noise  of high-expression \textit{E.~coli} genes is determined by the noise floor. \label{fig:transEff} \label{fig:noisefig} }
\end{figure}

\idea{Implications of growth robustness for translation.} Before continuing with the noise analysis, we to focus on the significance of the empirical relationship between the protein and message numbers (Eq.~\ref{eqn:empmod}).   
How can the cell counteract noise-induced reductions in robustness? Eq.~\ref{eqn:protmean} implies that gene expression can be thought of as a two-stage
 amplifier \cite{Crick:1970vy}: The first stage corresponds to transcription with a gain of message number $\mu_m$, and the second stage corresponds to translation with a gain in translation efficiency $\varepsilon$. (See Fig.~\ref{fig:transEff}AB.) The noise is completely determined by the first stage of amplification, provided that $\varepsilon \gg 0$ \cite{Paulsson:2000xi,Friedman:2006oh}. 
 Genes with low transcription levels are the noisiest. For these genes, the cell can achieve the same mean gene expression ($\mu_p$)  with lower noise by increasing the gain of the first stage (increasing message number) and decreasing the gain of the second stage (the translation efficiency) by the same factor. 
This is most clearly understood by reducing $\varepsilon$ at fixed $\mu_p$ in Eq.~\ref{eq:noiseTE}. 
 Highly transcribed genes have low noise and can therefore tolerate higher translation efficiency in the interest of economy (decreasing the total number of messages) \cite{Hausser:2019fi}. 
Growth robustness therefore predicts that the translation efficiency should grow with transcription level.

\idea{Translation efficiency increases with expression level in yeast.} The translation efficiency (Eq.~\ref{eqn:TE}) can be determined from the empirical translation model (Eq.~\ref{eqn:empmod}):
\begin{eqnarray} 
\varepsilon &=& 8.0\  \mu_m^{1.1},
\label{eqn:emptranseff}
\end{eqnarray}
as a function of message number. (An error analysis for both model parameters is described in Supplementary Material Sec.~\ref{sec:empmodel2}.)
In yeast, the translation efficiency clearly has a strong dependence on message number $\mu_m$, and grows with the expression level, exactly as predicted by robustness arguments. We note the contrast to the translation efficiency in \textit{E.~coli}, which  is roughly constant \cite{Balakrishnan:2022ai}. (See Supplementary Material Sec.~\ref{sec:noiseEcoli}.) We will speculate about the rationale for these differences in the discussion below.

\idea{Implications of growth robustness for transcription.} In addition to the prediction of translation efficiency depending on transcription, a second qualitative prediction of growth robustness is that essential gene expression should have a noise ceiling, or maximum noise level (Eq.~\ref{eq:noiseceil}), where noise above this level would be too great for robust growth. The fit between the statistical model and the observed noise has an important implication beyond confirming the predictions of the telegraph and statistical models for noise: The identification of the message number, $\mu_m$, as the key determinant of noise allows us to use this quantity as a proxy for noise in quantitative transcriptome analysis.


To identify a putative transcriptional floor, we now broaden our consideration beyond yeast to characterize the central dogma in two other model organisms: the bacterium \textit{Escherichia coli} and \textit{Homo sapiens} (human). We will also analyze three different transcriptional statistics for each gene: transcription rate ($\beta_m$), cellular message number ($\mu_{m/c}$), and message number ($\mu_m$). Analysis of these organisms explores orders-of-magnitude differences in characteristics of the central dogma, including total message number, protein number, doubling time,  message lifetime, and number of essential genes. (See Tab.~\ref{tab1}.) In particular, as a consequence of these differences, the three statistics describing transcription: transcription rate, cellular message number and message number are all distinct. Genes with matching message numbers in two different organisms will not have matching transcription rates or cellular message numbers. We hypothesize that cells must express essential genes above some threshold message number for robust growth; however, we expect to see that non-essential genes can be expressed at much lower levels since growth is not strictly dependent on their expression.
The signature of a noise-robustness mechanism would be the absence of essential genes for low message numbers. 


\begin{figure}[ht]
  \centering
   \includegraphics[width=0.45\textwidth]{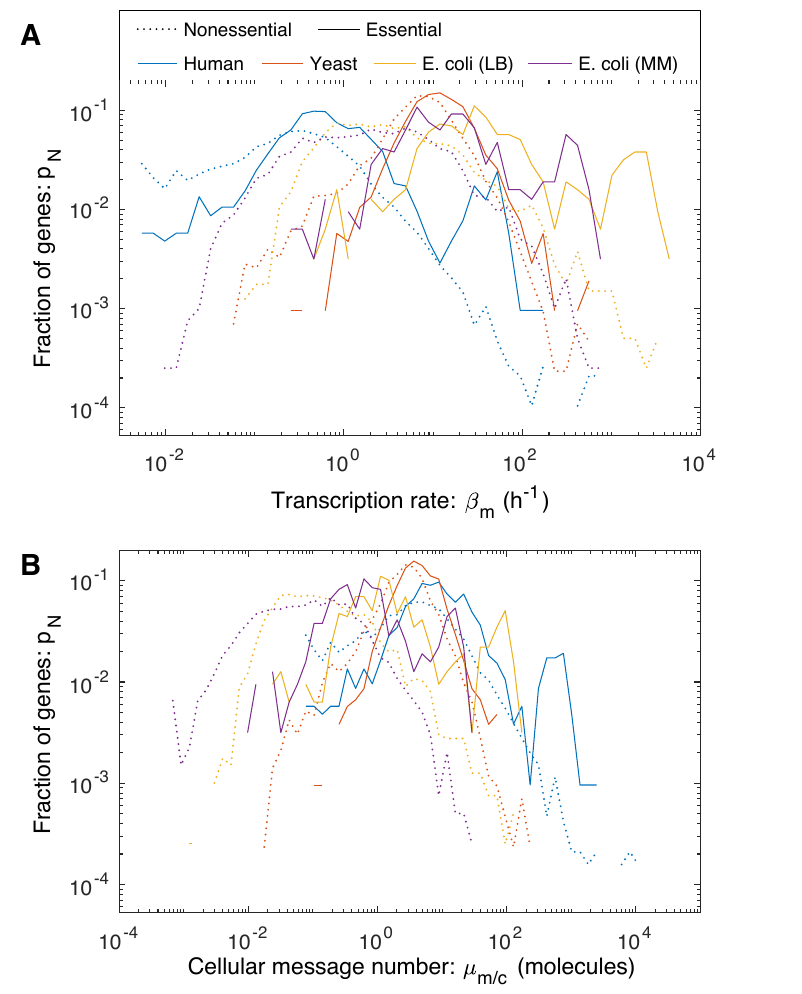}
      \caption{\textbf{Transcription in three model organisms.}  We characterized different gene transcriptional statistics in three model organisms. In \textit{E.~coli}, two growth conditions were analyzed. \textbf{Panel A: The distribution of gene transcription rate.} The transcription rate varies by two orders-of-magnitude between organisms.  \textbf{Panel B: The distribution of gene cellular message number.} There is also a two-order-of-magnitude variation between cellular message numbers.  
            \label{fig:wallAB}}
\end{figure}

\begin{figure*}[ht]
  \centering
   \includegraphics[width=0.99\textwidth]{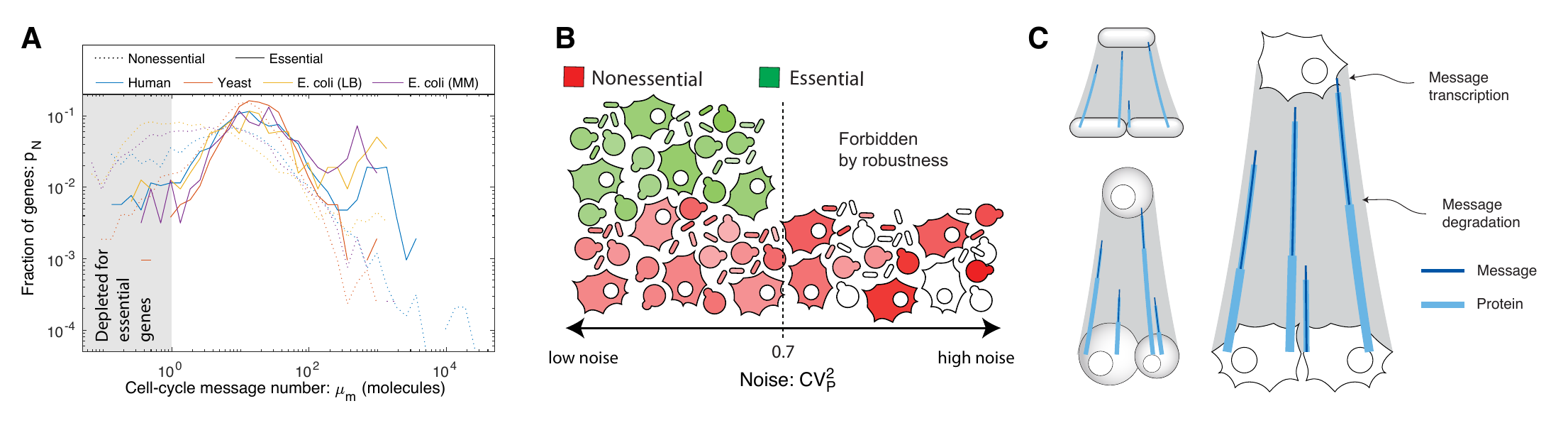}
      \caption{\textbf{Panel A: Transcription in three model organisms. The distribution of gene message number.}  All organisms have roughly similar distributions of message number for essential genes, which are not observed for message numbers below a couple per cell cycle. However, non-essential genes can be expressed at much lower levels.  \textbf{Panel B: Nonessential genes tolerate higher noise levels than essential genes.}  The floor of message number  is consistent with a noise ceiling of ${\rm CV}_p^2 = 0.7$ for essential genes (green). Nonessential genes  (red) are observed with lower transcription levels.  \textbf{Panel C: Conserved transcriptional program for essential genes.} The message number per gene (number of messages transcribed per cell cycle) is roughly identical in \textit{E.~coli}, yeast, and human. We show this schematically.  \label{fig:wallC}}
\end{figure*}



\begin{table*}
\resizebox{ \textwidth}{!}{\begin{tabularx}{ \textwidth}{ X  | X  X  X  X  X}
                            &                              \multicolumn{5}{c}{Estimated minimum essential gene} \cr
\centering{Model organism}  &   \centering{Maximum essential gene noise:}         &  \centering{ messages /cell-cycle: }   & \centering{messages /cell: } & \centering{transcription rate: } & \centering{proteins:} \cr
                         &       \centering{ $\max{\rm CV}_p^2$}     &          \centering{$\mu_{m}^{\rm min}$ }     & \centering{$\mu_{m/c}^{\rm min}$ } & \centering{$\beta_{m}^{\rm min}$ (h$^{-1}$) } & \centering{$\mu_p^{\rm min}$} \cr
              
\hline
\hline
 \textit{E.~coli} (LB) & \raggedleft   \centering{$0.7$}  &    \raggedleft   \centering{$1$} & \raggedleft   \centering{$0.08$}  & \raggedleft \centering{$2$}   & \raggedleft  \centering{$30$}  \cr 
  \ \ \ \ \ \ \ \ \ \ \ (M9) &   \raggedleft  \centering{ $0.7$}  &  \raggedleft   \centering{$1$}  & \raggedleft   \centering{$0.03$}  & \raggedleft \centering{$0.7$}  & \raggedleft \centering{$30$}  \cr 
                     \hline
Yeast                 & \raggedleft   \centering{$0.7$}  & \raggedleft \centering{$1$}  & \raggedleft \centering{$0.2$}   & \raggedleft \centering{$0.7$}  & \raggedleft  \centering{$400$}  \cr 
                                          \hline
Human                      & \raggedleft   \centering{$0.7$}  & \raggedleft  \centering{$1$} & \raggedleft \centering{$0.6$}  & \raggedleft \centering{$0.04$} & \raggedleft \centering{$3000$}  \cr            
                     \hline
\end{tabularx}}
\caption{ \textbf{Estimates of threshold levels for the central dogma in three model organisms.} Estimates for the lower thresholds of transcription statistics as inferred from our analysis based on the \textit{one-message-per-cell-cycle rule}. 
\label{tab2} }
\end{table*}

\idea{No organism-independent threshold is observed for transcription rate or cellular message number.} 
Histograms of the per-gene transcription rate and cellular message number are shown in Fig.~\ref{fig:wallAB} for \textit{E.~coli}, yeast, and human. Consistent with existing reports, essential genes have higher expression than non-essential genes on average; however, there does not appear to be any consistent threshold in \textit{E.~coli} (even between growth conditions), yeast, or human transcription, either as characterized by the transcription rate ($\beta_m$) or the cellular message number ($\mu_{m/c}$). For instance, the per gene rate of transcription is much lower in human cells than \textit{E.~coli} under rapid growth conditions, with yeast falling  in between.

\idea{An organism-independent threshold is observed for message number for essential genes.} In contrast to the other two transcriptional statistics, there is a consistent lower limit, or floor, on message number ($\mu_m$) of somewhere between 1 and 10 messages per cell cycle for essential genes. (See Fig.~\ref{fig:wallC}.) Non-essential genes can be expressed at a much lower level. This floor is consistent not only between \textit{E.~coli}, growing under two different conditions, but also between the three highly-divergent organisms: \textit{E.~coli}, yeast and human.  We will conservatively define the minimum message number as 
\begin{equation}
\mu_m^{\rm min} \equiv 1,    
\end{equation} 
and summarize this observation as the \textit{one-message-per-cell-cycle rule} for essential gene expression.  

In addition to the common floor for essential genes, there is a common gene expression distribution shape shared between organisms dependent on the message numbers, especially for low-expression essential genes.  This is observed in spite of the significantly larger number of essential genes in human relative to \textit{E.~coli}. (See Fig.~\ref{fig:wallC}.) Interestingly, there is also a similarity between the non-essential gene distributions for \textit{E.~coli} and human, but not for yeast, which appears to have a much lower fraction of genes expressed at the lowest message numbers.

\idea{What genes fall below-threshold?} We have hypothesized that essential genes should be expressed above a threshold value for robustness. It is therefore interesting to consider the function of genes that fall below this proposed threshold. Do functions of these genes give us any insight into essential processes that do not require robust gene expression?

Since our own preferred model system is \textit{E.~coli}, we focus here. Our essential gene classification was based on the construction of the Keio knockout library \cite{Baba:2008vn}. By this classification, 10 essential genes were below threshold. (See Supplementary Material Tab.~\ref{ess_tab}.) Our first step was to determine what fraction of these genes were also classified as essential using transposon-based mutagenesis \cite{Gerdes:2003ty,Goodall:2018rm}. Of the 10 initial candidates, only one gene, \textit{ymfK}, was consistently classified as an essential gene in all three studies, and we estimate that its message number is just below the threshold ($\mu_m = 0.4$). \textit{ymfK} is located in the lambdoid prophage element e14 and is annotated as a CI-like  repressor which regulates lysis-lysogeny decision \cite{Mehta:2004sy}. In $\lambda$ phase, the CI repressor represses lytic genes to maintain the lysogenic state. A conserved function for \textit{ymfK} is consistent with it being classified as essential, since its regulation would prevent cell lysis.
However, since \textit{ymfK} is a prophage gene, not a host gene, it is not clear that its expression should optimize host fitness, potentially at the expense of phage fitness. 
In summary, closer inspection of  below-threshold essential genes supports the threshold hypothesis.

\idea{Maximum noise for essential genes.} The motivation for hypothesizing a minimum threshold for message number was noise-robustness, or the existence of a hypothesized  noise ceiling above which essential gene expression is too noisy to allow robust cellular proliferation. With the \textit{one-message-per-cell-cycle rule}, $\mu_m^{min} \equiv 1$, we can estimate the essential gene noise ceiling using Eq.~\ref{eqn:protnoise}:
\begin{equation}
{\rm CV}_p^2 \le 0.7, \label{eqn:noiseceiling}
\end{equation}
for essential genes. 
Since noise depends only on the message number, we expect to observe the same limit in all organisms if the message number floor is conserved.

\idea{Estimating the floor on central-dogma parameters.} If message number floor is conserved, a limit can be estimated for the floor value on other transcriptional parameters. Using Eq.~\ref{eqn:cellularmn}, we can estimate the floor on the cellular message number (as measured in RNA-Seq measurements):  
\begin{equation}
\mu_{m/c}^{\rm min} = \textstyle\frac{\tau_{m} }{T}, 
\end{equation}
for essential genes. 
Similarly, we can use Eq.~\ref{eqn:shape2} to estimate the minimum transcription rate: 
\begin{equation}
\beta_m^{\rm min} = \textstyle\frac{1}{T}, \label{eqn:transmin}
\end{equation}
for essential genes. 
Again, this result has an intuitive interpretation as the one-message-per-cell-cycle rule.  Finally, we can estimate a floor on essential protein abundance, assuming a constant translation efficiency using Eq.~\ref{eqn:protmean}:
\begin{equation}
\mu_p^{\rm min} = {\varepsilon}, 
\end{equation}
for essential genes, where $\varepsilon$ is the translation efficiency (which we will assume is well approximated by the mean in the context of the estimate). All four floor estimates for each model organism are shown in Tab.~\ref{tab2}.

\section*{Discussion}

\idea{Noise by the numbers.} 
Although there has already been significant discussion  of the scaling of biological noise with protein abundance \cite{Paulsson:2000xi,Friedman:2006oh,Bar-Even:2006rv,Newman:2006nl,Taniguchi2010}, our study is arguably the first to test the predictions of the telegraph and statistical noise models against absolute measurements of protein and message abundances. This approach is particularly important for the message number ($\mu_m$), which determines the magnitude of the noise in protein expression, and facilitates direct comparisons of noise between organisms as well as identifying the common distributions of message number for genes, that are conserved from bacteria to human.

\idea{Noise scaling in \textit{E.~coli} versus yeast.} A key piece of evidence for the significance of the message number was the observation of the non-canonical scaling of the yeast noise with protein abundance (Fig.~\ref{fig:noisefig}); however, the canonical model (Eq.~\ref{eqn:nullmodel}) does accurately describe the noise  in \textit{E.~coli} (see Fig. \ref{fig:EcoliNoise3param}). Why does the noise scale differently? In \textit{E.~coli}, the translation efficiency is only weakly correlated with the gene expression \cite{Balakrishnan:2022ai}, and therefore the canonical model is a reasonable approximation (Supplementary Material Sec.~\ref{sec:yeastnoiseempiricalmodel}).  However, we also argued that translation efficiency should grow with expression level.  Why is this not observed in \textit{E.~coli}? Due to the high noise floor in \textit{E.~coli}, nearly all essential genes are expressed at a sufficiently high expression level such that the noise is dominated by the noise floor \cite{Taniguchi2010}. As a consequence, increasing the message number, while decreasing translation efficiency, does not decrease the noise even as it increases the metabolic load as a result of increased transcription. (A closely related point has recently been made in \textit{Bacillus subtilis} \cite{Deloupy:2020oj}, where Deloupy \textit{et al.}~report that the noise cannot be tuned by adjusting the message number due to the noise floor.) Our expectation is therefore that other bacterial cells will look similar to \textit{E.~coli}: They will have a higher noise floor and a similar scaling of noise with protein abundance. 

In contrast, due to the lower noise floor, we expect eukaryotic cells to optimize the central dogma processes like yeast and as a result will have a similar non-canonical scaling of noise with protein abundance. Although this non-canonical scaling is clear from the abundance data (Fig.~\ref{fig:noisefig}B), there is an important qualification to emphasize: the mechanism that gives rise to the non-canonical scaling is due to the correlation between translation efficiency and transcription. Regulatory changes that effect only transcription (\textit{i.e.}~increase $\mu_m$) and not translation ($\varepsilon$) should obey the canonical noise model (Eq.~\ref{eqn:nullmodel}). This scenario may help explain why Bar-Even \textit{et al.}~claim to observe canonical noise scaling in yeast \cite{Bar-Even:2006rv}, studying a subset of genes under a range of conditions resulting in differential expression levels. The failure of the canonical noise model (Eq.~\ref{eqn:nullmodel}) at the proteome level in yeast (Eq.~\ref{eqn:noisemodelstat}) is a consequence of genome-wide optimization of the relative transcription and translation rates. 

\idea{Essential versus non-essential genes.} What genes are defined as \textit{essential} is highly context specific \cite{Chin:2012dd}. It is therefore important to consider whether the comparison between these two classes of genes is informative in the context of our analysis. We believe the example of \textit{lac} operon in \textit{E.~coli} is particularly informative in this respect. The genes \textit{lacZYA} are conditionally essential: they are required when lactose is the carbon source; however, these genes are repressed when glucose is the carbon source.  Our expectation is that these conditionally essential genes will obey the one-message-per-cell-cycle rule when these genes are required; however, they need not obey this rule when the genes are repressed.  By analyzing  essential genes, we are limiting the analysis to transcriptionally-active genes, whereas the non-essential category contains both transcriptionally-active and silenced genes.


\idea{Protein degradation and transcriptional bursting.} Two important mechanisms can act to significantly increase the noise above the levels we predict: protein degradation and transcriptional bursting. Although the dominant mechanism of protein depletion is dilution in \textit{E.~coli}, protein degradation plays an important role in many organisms, especially in eukaryotic cells \cite{Ciechanover:2005vj,Eden2011}. If protein degradation depletes proteins faster than dilution, the shape parameter decreases below our estimate (Eq.~\ref{eqn:shape2}), increasing the noise. Likewise, the existence of transcriptional bursting, in which the chromatin switches between transcriptionally active and quiescent periods, can also act to increase the noise \cite{Raser:2005we,Golding:2005ok,Iyer-Biswas:2009sd}. Since the presence of both these mechanisms increases the noise beyond what is predicted by the message number, they do not affect our estimate of the minimum threshold for $\mu_m$.

\idea{The biological implications of noise.} What are the biological implications of gene expression noise? Many important proposals have been made, including bet-hedging strategies, the necessity of feedback in gene regulatory networks, \textit{etc} \cite{Raser:2005we}. Our analysis suggests that noise influences the optimal function of the central dogma process generically. Hausser \textit{et al.} have already discussed some aspects of this problem and use this approach to place coarse limits on transcription versus translation rates \cite{Hausser:2019fi}. The transcriptional floor for essential genes that we have proposed places much stronger limits on the function of the central dogma. 

Although we describe our observations as a floor, a more nuanced description of the phenomenon is a common distribution of gene message numbers, peaked at roughly 15 messages per cell cycle and cutting off close to one message per cell cycle. Does this correspond to a hard limit? We expect that this does not since there are a small fraction of genes, classified as essential, just below this limit; however, it does appear that virtually all essential genes have optimal expression levels above this threshold.
The common distribution of message number clearly suggests that noise considerations shape the  function of the central dogma for virtually all genes. 
Exploring this hypothesis will require quantitative models that explicitly realize the high cost of noise-induced low essential-protein abundance. We will present such an analysis elsewhere.

\idea{Adapting the central dogma to increased cell size and complexity.}
Although core components of the central dogma machinery are highly-conserved, there has been significant complexification of  both the transcriptional and translational processes in eukaryotic cells  \cite{Cooper:2000af}.  Given this increased regulatory complexity, it is unclear how the central dogma processes should be adapted in larger and more complex cells. An important clue to this adaptation comes from \textit{E.~coli} proliferating with different growth rates. Although there are very significant differences between the cellular message number as well as the overall transcription rate under the two growth conditions, there is very little difference in message number. In short, roughly the same number of messages are made during the cell cycle, but they are made more slowly under slow growth conditions. 

How does this picture generalize in eukaryotic cells? Although both the total number of messages and the number of essential and non-essential genes are larger in both yeast and human cells, the distribution of the message number per gene is essentially the same as \textit{E.~coli} (Fig.~\ref{fig:wallC}).  The  conservation of the message number between organisms is consistent with all of these organisms being optimized with respect to the same trade-off between economy and robustness to noise.

\idea{Data availability.} We include a source data file which includes the estimated message numbers as well as essential/nonessential classifications for each organism.
 
\idea{Acknowledgments.} The authors would like to thank B.~Traxler, A.~Nourmohammad, J.~Mougous, K.~Cutler, M.~Cosentino-Lagomarsino, S.~van Teeffelen, and S.~Murray. This work was supported by NIH grant R01-GM128191.

\idea{Author contributions:} T.W.L., H.K.J.C., D.H. and P.A.W.~conceived the research. T.W.L. and P.A.W. performed the analysis. H.K.J.C. and D.H. performed experiments and analysis. T.W.L., H.K.J.C., D.H. and P.A.W. wrote the paper.

\idea{Competing interests:}
The authors declare no competing interests.

\bibliographystyle{naturemag}
\bibliography{message}

\onecolumngrid

\newpage

\appendix

\tableofcontents

\twocolumngrid

\section{Supplemental analysis}

\subsection{Gillespie Simulation of the telegraph model}
\label{SecGS}


Protein distributions of the telegraph model for \textit{E.~coli} were simulated with a Gillespie algorithm. Assuming the lifetime of the cell cycle ($T_{cc} = 30$ min) \cite{Bernstein:2002rp}, mRNA lifetime ($\tau_m = 2.5$ min)  \cite{Chen:2015wt}, and translation rate ($\beta_p \approx 500$ hr$^{-1}$), the protein distributions for several mean expression levels were numerically generated for exponential growth with 100,000 stochastic cell divisions, with protein partitioned at division following the binomial distribution. 

The gamma distributions for each mean message number with scale and shape parameters determined by the corresponding translation efficiency and message number  ($\theta = \varepsilon \ln 2 $, $k = \frac{\mu_m}{\ln 2} $) as used for the Gillespie simulation were also plotted with the protein distributions.

\begin{equation}
    p(n | \theta, k) = \frac{1}{\Gamma(k) \theta^k} n^{k-1} e^{-\frac{n}{\theta}} 
\end{equation}

\begin{figure}
    \centering
    \includegraphics[width=0.45\textwidth]{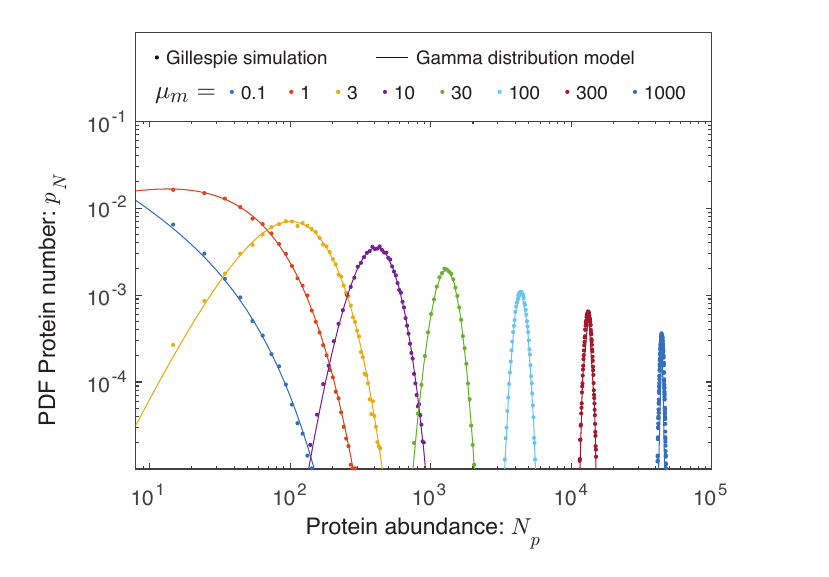}
    \caption{\textbf{The protein abundance is approximately gamma distributed.} Protein abundance was modeled for eight different transcription rates using a Gillespie simulation, including the stochastic partitioning of the proteins between daughter cells at cell division. The range in abundance matches the observed range of expression levels in the cell. 
      We observed that the simulated protein abundances were well fit by gamma distributions.  }
    \label{fig:gillespie}
\end{figure}

\begin{table*}
\resizebox{\textwidth}{!}{\begin{tabularx}{1.15 \textwidth}{ X | X | X  X X | X  X  X | X X }
                         &            &                   &                             &                               &  \multicolumn{3}{c|}{Total number of }       &    \multicolumn{2}{c}{Average}      \cr
\centering{Model organism}                &  \centering{Growth condition} & \centering{Doubling time}:  & \centering{Message lifetime}: &  \centering{Message recycling ratio:} & \centering{messages /cell:}      &   \centering{messages /cell-cycle:}          &    \centering{proteins:}     &  \centering{translation efficiency:}  &  \centering{translation rate:} \cr
                         &                               & \centering{$T$}    &  \centering{$\tau_m=\gamma_m^{-1}$} & \centering{ $m=T/\tau_m$ }             &   \centering{$N^{\rm tot}_{m/c}$}  &   \centering{$N^{\rm tot}_{m}$}    &  \centering{$N^{\rm tot}_{p}$}   & \centering{$\varepsilon$}  &  \centering{$\beta_p$ (h$^{-1}$)}  \cr
\hline
\hline
 \textit{Escherichia coli}  & \raggedleft   LB & \raggedleft  30 min \cite{Bernstein:2002rp}  & \raggedleft $2.5$ min \cite{Chen:2015wt} & \raggedleft 12  & \raggedleft $7.8 \times 10^3$ \cite{Bartholomaus:2016df}   & \raggedleft $9.4 \times 10^4$ & \raggedleft $3 \times 10^6$ \cite{Milo:2013yn} & \raggedleft  $22$  & \raggedleft  $530$   \cr 
 (\textit{E.~coli})         & \raggedleft   M9 & \raggedleft  90 min \cite{Bernstein:2002rp} & \raggedleft $2.5$ min  \cite{Chen:2015wt} & 36 \raggedleft & \raggedleft $2.4 \times 10^3$ \cite{Bartholomaus:2016df} & \raggedleft $8.6 \times 10^4$ & \raggedleft $3 \times 10^6$ \cite{Milo:2013yn}  & \raggedleft $24$ & \raggedleft  $580$  \cr 
                     \hline
 \textit{Sacchromyces cerevisiae (Yeast--haploid)}     & \raggedleft  YEPD    & \raggedleft  90 min \cite{Albe_2002_book} & \raggedleft  22 min \cite{Chia:1979jm} & \raggedleft 4 & \raggedleft $2.9 \times 10^4$ \cite{Pelechano:2010gz} & \raggedleft $1 \times 10^5$ & \raggedleft  $5 \times 10^7$ \cite{Futcher:1999sb} & \raggedleft $4 \times 10^2$   & \raggedleft $410$  \cr 
                                          \hline
 \textit{Homo sapiens (Human)}         & \raggedleft   Tissue   & \raggedleft  24 h  \cite{Cooper:2000af}  & \raggedleft  14 h  \cite{Yang:2003mf} & \raggedleft  1.7 & \raggedleft   $3.6 \times 10^5$ \cite{Albe_2002_book} & \raggedleft $5 \times 10^5 $  & \raggedleft $2 \times 10^9$ \cite{Milo:2013yn} & \raggedleft  $4 \times 10^3$ & \raggedleft  120  \cr 
                     \hline
\end{tabularx}}

\caption{ \textbf{Central dogma parameters for three model organisms with detailed references.}  Columns three through seven hold representative values for  measured central-dogma parameters for the model organisms described in the paper. Each value is followed by a reference for its source. The last three columns hold estimates for the lower threshold on transcription inferred from our analysis. \label{tab1refs} The sources of the numbers and estimates are described in the Supplemental Material Sec.~\ref{sec:desriptonTab1}.}
\end{table*}

\subsection{ Selection of  central dogma parameter estimates}
\label{sec:desriptonTab1}

The estimates for central dogma model parameters come from two types of data: (i) quantitative measurement of cellular-scale parameters for each organism (total number of messages in the cell, cell cycle duration, \textit{etc}) and (ii) genome-wide studies quantitative of mRNA and protein abundance.

For the cellular-scale central dogma parameters, we relied heavily on an online compilation of biological numbers: BioNumbers \cite{Milo:2010nw}. This resource provides a collection of curated quantitative estimates for biological numbers, as well as their original source. In the interest of conciseness, we have cited only the original source in the Tab.~\ref{tab1}, although we are extremely grateful and supportive of the creators of the BioNumbers website for helping us very efficiently identify consensus estimates for the parameters of the central dogma parameters.

For the selection of genome-wide studies on abundance, we used many of the same resources cited in BioNumbers as well as studies selected by a previous study of a quantitative analysis of the central dogma: Hausser \textit{et al.} \cite{Hausser:2019fi}.

\subsubsection{\textit{E.~coli} data}

\idea{Message lifetimes:} The message lifetimes (and median lifetime) were taken from a recent transcriptome-wide study by Chen \textit{et al.}  \cite{Chen:2015wt}. These investigators measured the lifetime in both rapid (LB) and slow growth (M9). 

\idea{Noise:} Taniguchi \textit{et al.} 
have performed a beautiful  simultaneous study of the proteome and transcriptome with single-molecule sensitivity \cite{Taniguchi2010}. Although we use the noise analysis data from this study for our supplemental analysis of \textit{E.~coli} noise, it is not the source for our \textit{E.~coli} transcriptome data due to the extremely slow growth of the cells in this study (150 minute doubling time), which is not consistent with the growth conditions for the other sources of data. 

\idea{mRNA abundance:} Instead, we used data from the more recent Bartholomaus \textit{et al.} study \cite{Bartholomaus:2016df}, which characterizes the transcriptome in both rapid (LB) and slow growth (M9). 

\idea{Total cellular message number.} This study was chosen since it was the source of the BioNumbers estimates of cellular message number in \textit{E.~coli} (BNID 112795 \cite{Milo:2010nw}). 

\idea{Doubling time:} The source of the doubling times for rapid (LB) and slow (M9) growth of \textit{E.~coli} comes from Bernstein \cite{Bernstein:2002rp}. 

\idea{Essential gene classification.} The classification of essential genes in yeast comes from the construction of the Keio knockout collection from Baba \textit{et al.} \cite{Baba:2008vn}.

\idea{Protein number.} The total protein number in \textit{E.~coli} came from Milo's recent review of this subject \cite{Milo:2013yn}.

\subsubsection{Yeast data}

\idea{Message lifetimes:} The message lifetimes (and median lifetime) were taken from Chia \textit{et al.} \cite{Chia:1979jm}.

\idea{Noise:} The noise data was taken from the Newman \textit{et al.} study, which used flow cytometry of a library of fluorescent fusions to characterize protein abundance with single-cell resolution \cite{Newman:2006nl}.

\idea{mRNA abundance:} The transcriptome data comes from the very recent Blevins \textit{et al.} study \cite{Blevins:2019yj}.

\idea{Total cellular message number.} There are a wide-range of estimates for the total cellular message number in yeast: $1.5\times 10^4$ \cite{Hereford:1977au} (BNID 104312 \cite{Milo:2010nw}), $1.2\times10^4$ \cite{Haar:2008eg} (BNID 102988 \cite{Milo:2010nw}), $6.0\times 10^4$ \cite{Zenklusen:2008fo} (BNID 103023 \cite{Milo:2010nw}), $2.6\times 10^4$ \cite{Pelechano:2010gz} (BNID 106763 \cite{Milo:2010nw}) and $3.0\times 10^4$ \cite{Miura:2008zn}. We used the compromise value of $2.9\times 10^4$.

\idea{Doubling time:}  The doubling time was taken from \cite{Albe_2002_book}.

\idea{Protein number.} The total protein number in yeast comes from Futcher \textit{et al.} \cite{Futcher:1999sb}.

\idea{Essential gene classification.} The classification of essential genes in yeast comes from van Leeuwen \textit{et al.} \cite{Leeuwen:2020eh}.

\idea{Proteome abundance data:} The proteome abundance data came from two sources: flow cytometry of fluorescent fusions from Newman \textit{et al.} \cite{Newman:2006nl} as well as mass-spec data from de Godoy \textit{et al.} \cite{Godoy:2008eb}.

\subsubsection{Human data}

\idea{Message lifetimes:} The message lifetimes (and median lifetime) were taken from Yang \textit{et al.} \cite{Yang:2003mf} who reported a median half life of 10 h which corresponds to a lifetime of 14 h.

\idea{mRNA abundance:} The transcriptome data comes from the data compiled by the Human Protein Atlas \cite{Uhlen:2015jm}, which we averaged over tissue types.

\idea{Total cellular message number.} The total cellular message number in human comes from Velculescu \textit{et al.} \cite{Velculescu:1999hf} (BNID 104330 \cite{Milo:2010nw}).

\idea{Doubling time:}  The doubling time was taken from \cite{Cooper:2000af}.

\idea{Protein number.} The total protein number in human came from Milo's recent review of this subject \cite{Milo:2013yn}.

\idea{Essential gene classification.} The classification of essential genes in human comes from Wang \textit{et al.} \cite{Wang:2015lb}.

\subsection{Quantitative estimates of central dogma parameters}

\subsubsection{Estimating the \textit{cellular message number}: $\mu_{m/c}$}

For each model organism (and condition), we found a consensus estimate from the literature for the total number of mRNA messages per cell $N_{m/c}^{\rm tot}$. This number and its source are provided in Tab.~\ref{tab1}. To estimate the number of messages corresponding to gene $i$, we re-scaled the un-normalized abundance level $r_i$:
\begin{equation}
    N_{m/c, i} = N_{m/c}^{\rm tot}  \textstyle\frac{r_i}{\sum_j r_j}, \label{eqn:rescalcmn}
\end{equation}
where the sum over gene index $j$ runs over all genes.

\subsubsection{Estimating the \textit{transcription rate}: $\beta_{m}$}
\label{sec:esttransrate}
To estimate the transcription rate for gene $i$, we start from the estimated cellular message number $N_{m/c,i}$ and use the telegraph model prediction for the cellular message number:
\begin{equation}
    N_{m/c,i} = \beta_{m,i}/\gamma_{m,i},
\end{equation}
where $\gamma_{m,i}$ is the message decay rate. Since gene-to-gene variation in message number is dominated by the transcription rate (\textit{e.g}~\cite{Chen:2015wt}), we estimate the decay rate as the inverse gene-median message life time:
\begin{equation}
\gamma_{m,i} = \tau_m^{-1}, 
\end{equation}
for which a consensus value was found from the literature. This number and its source are provided in Tab.~\ref{tab1}. We then estimate the gene-specific transcription rate:
\begin{equation}
    \beta_{m,i}=N_{m/c,i}/\tau_m.
\end{equation}

\subsubsection{Estimating the \textit{message number}:  $\mu_{m}$}
\label{sec:CCMN}

To estimate the message number of gene $i$, we use the predicted value from the telegraph model:
\begin{equation}
    N_{m,i} = T \beta_{m,i} = \textstyle \frac{T}{\tau_m} N_{m/c,i},
\end{equation}
where $T$ is the doubling time and $N_{m/c,i}$ is the cellular message number (Eq.~\ref{eqn:rescalcmn}).

\subsubsection{Estimating the \textit{protein number}: $\mu_p$ }
\label{sec:MS}
\label{sec:Fl}

\begin{figure}
    \centering
    \includegraphics[width=0.48\textwidth]{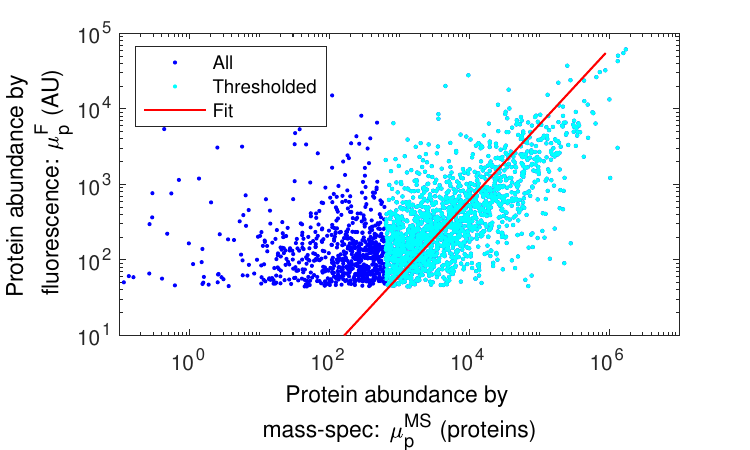}
    \caption{\textbf{ Fit to rescale fluorescence intensity to protein number. } Protein abundance from flow cytometry fluorescence \cite{Newman:2006nl} as a function of mass-spectrometry scaled abundance \cite{Godoy:2008eb}. The mass-spectrometry data was thresholded at 10 proteins, and then a linear fit was performed to find the offset of 3.9, which was used to convert protein fluorescence AU to number.}
    \label{fig:AUconv}
\end{figure}

The protein abundance data for yeast grown in YEPD media and measured with flow cytometry fluorescence \cite{Newman:2006nl} were given in arbitrary units (AU). In order to convert from AU to protein number, the fluorescence values were rescaled by comparing with mass-spectrometry protein abundance data for yeast grown in YNB media \cite{Godoy:2008eb}. Since the protein abundance from mass-spectrometry was given in terms of Intensity, the Intensity values were first rescaled by the total number of proteins in yeast, $5 \times 10^7$. The mass-spectrometry protein data was thresholded at 10 proteins, based on the assumption that the noise of the data for 10 and fewer proteins makes the data unreliable. Next, the log of the fluorescence protein abundance in AU as a function of the log of thresholded mass-spectrometry protein abundance was fit as a linear function with an assumed slope of 1 to find the offset, 3.9, (Fig. \ref{fig:AUconv}) which corresponds to a multiplicative scaling factor (Eqn. \ref{eqn:logconversion}). We then used that offset value to rescale the fluorescence data from AU to protein number. We also compared to yeast grown in SD media \cite{Newman:2006nl} and found a similar offset result.

\begin{eqnarray}
    \log \mu_P^{\text{F}} = m \log \mu_P^{\text{MS}} + b \\
    \mu_P^{\text{F}} = b(\mu_P^{\text{MS}})^m \label{eqn:logconversion}
\end{eqnarray}


\subsection{Empirical models for yeast gene expression}

To generate the empirical model for protein number as a function of message number, we used protein abundance data from Newman \textit{et al.} \cite{Newman:2006nl}, re-scaled to estimate protein number (Sec.~\ref{sec:Fl}) and transcriptome data from Lahtvee \textit{et al.} \cite{Lahtvee:2017al}, re-scaled to estimate message number (Sec.~\ref{sec:CCMN}).

\subsubsection{The meaning of the error estimates}

Before providing a detailed error analysis, it is important to place our error estimates in perspective. The error that we will be estimating is the statistical error associated with the finite sample size; however, \textit{this is not the dominant source of error}. A far more important consideration are systematic problems with our analysis and the underlying experiments. For instance, since we do not have a detailed model for the error of the experiments analyzed, there are multiple distinct analyses (\textit{i.e.}~assumptions about the error model) that could be implemented for the data fitting, each giving slightly different model parameters. These model to model differences still give rise to predictions consistent with our qualitative conclusions; however, they are likely larger than the statistical uncertainty we compute (while assuming a particular model).

\subsubsection{Empirical model for protein number}

\label{sec:empmodel0}
We initially fit the empirical model for protein number,
\begin{equation}
    \mu_p = C_0 \mu_m^{\alpha_0}, \label{eqn:pn}
\end{equation}
to the data using a standard least-squares approach; however, the algorithm led to a very poor fit since it does not account for uncertainty in both independent and dependent variables. We therefore used an alternative approach \cite{Hellton2014}, which assumes comparable error in both variables. The model parameters are: 
\begin{eqnarray}
\alpha_0 &=& 2.1\pm 0.04,\\
C_0 &=& 8.0 \pm 1.0,
\end{eqnarray}
where the uncertainties are the estimated standard errors.

\subsubsection{Empirical model for message number}

For the prediction of the coefficient of variation, it is useful to invert Eq.~\ref{eqn:pn} to generate a model for message number as a function of protein number:
\begin{eqnarray}
    \mu_m &=& C_{0}^{-1/\alpha_{0}} \mu_p^{1/\alpha{0}}, \\
          &=& C_1 \mu_p^{\alpha_1}, \label{eqn:ccmn}
\end{eqnarray}
where the last line defines two new parameters: a coefficient $C_1$ and an exponent $\alpha_1$. The resulting parameters and uncertainties are: 
\begin{eqnarray}
\alpha_1 &\equiv& 1/\alpha_0,\\
&=& 0.48\pm 0.01,\\
C_1 &\equiv& C_0^{-1/\alpha_0},\\
&=& 0.37\pm 0.02,
\end{eqnarray}
where the uncertainties are the estimated standard errors.

\subsubsection{Empirical model for translation efficiency}

\label{sec:empmodel2}

To generate an empirical model for translation efficiency, we started from the empirical model for protein number (Eq.~\ref{eqn:pn}),  and then use Eq.~\ref{eqn:protmean} to relate protein number, message number, and translation efficiency: 
\begin{eqnarray}
        \varepsilon &=& \textstyle \frac{\mu_p}{\mu_m},\\ &=& C_0 \mu_m^{\alpha_0-1}, \\
        &=& C_2 \mu_m^{\alpha_2},
\end{eqnarray}
where the last line defines two new parameters: a coefficient $C_2$ and an exponent $\alpha_2$. The resulting parameters and uncertainties are: 
\begin{eqnarray}
\alpha_2 &=& \alpha_0-1,\\
&=& 1.07 \pm 0.04,\\
C_2 &=& C_0,\\
&=& 8.0\pm 1.0,
\end{eqnarray}
where the uncertainties are the estimated standard errors.

\subsubsection{Empirical model for the coefficient of variation}

\label{sec:empmodel3}

To generate an empirical model for the coefficient of variation, we started from the empirical model for message number (Eq.~\ref{eqn:ccmn}),  and then substitute this into the statistical model prediction for CV$_p^2$ (Eq.~\ref{eqn:protnoise}):
\begin{eqnarray}
{\rm CV}_p^2 &=& \textstyle \frac{\log 2}{\mu_m},\\
 &=&  C_0^{1/\alpha_0}\log 2 \cdot \mu_p^{-1/\alpha_0},\\
 &=& C_3 \mu_p^{\alpha_3}, 
\end{eqnarray}
where the last line defines two new parameters: a coefficient $C_3$ and an exponent $\alpha_3$. The resulting parameters and uncertainties are: 
\begin{eqnarray}
\alpha_3 &\equiv& -1/\alpha_0,\\
&=& -0.48\pm0.01, \\
C_3 &\equiv& C^{1/\alpha_0}_0\log 2,\\
&=& 1.9\pm 0.1,
\end{eqnarray}
where the uncertainties are the estimated standard errors.

\subsection{Supplemental analysis of gene expression noise}
\label{sec:SMnoise} 

The quantitative model for gene expression noise includes multiple contributions: 
\begin{equation}
    {\rm CV}^2_p \approx \textstyle\frac{1}{\mu_p} + \textstyle\frac{\log 2}{\mu_m} + c_0, 
\end{equation}
where the first term can be understood to represent the Poisson noise from translation, the second term the Poisson noise from transcription, and the last term, $c_0$, is called the \textit{noise floor} and is believed to be caused by the cell-to-cell variation in metabolites, ribosomes, and polymerases \textit{etc} \cite{Elowitz:2002tb,Swain:2002te}.

\begin{figure}
    \centering
    \includegraphics[width=0.48\textwidth]{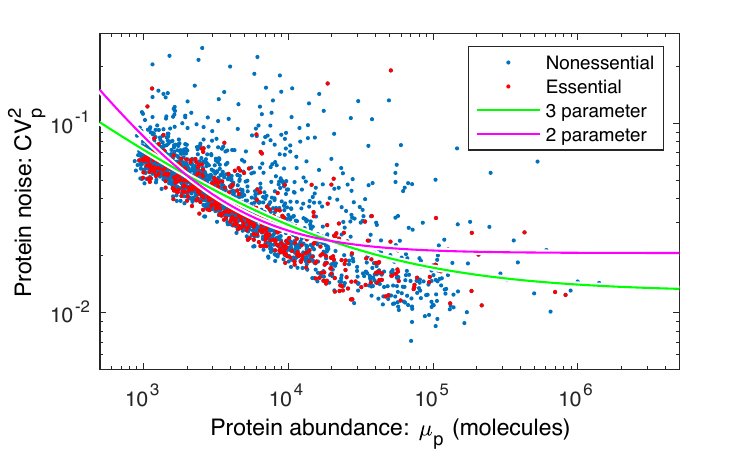}
    \caption{\textbf{ Yeast noise fit against canonical noise model, with a noise floor. } Yeast noise data fit with the 2- (null hypothesis with $\mu_p^{-1}$ dependence) and 3- parameter ($\mu_p^a$) models. }
    \label{fig:yeastnoise}
\end{figure}

\subsubsection{Inclusion of noise floor in the yeast analysis}

 In the main text of the paper, we have ignored the role of the noise floor in the analysis of noise in yeast. Unlike \textit{E.~coli}, where the noise floor is high (${\rm CV}^2_p=0.1$) and is determinative of the noise associated with almost all essential genes \cite{Taniguchi2010,Elowitz:2002tb,Swain:2002te}, in yeast the noise floor is much lower (${\rm CV}^2_p=0.01$) and therefore affects only genes with the highest expression.  
 
 In this section, we will consider models that include the noise floor, since its presence can make the noise scaling more difficult to interpret. To determine if the scaling of the noise is consistent with the canonical assumption that the noise is proportional to $\mu_p^{-1}$ for low expression, we  will consider two competing empirical models for the noise (Fig. \ref{fig:yeastnoise}). In the null hypothesis, we will consider a model:
 \begin{equation}
 \eta_0(\mu_p; b,c ) = \textstyle\frac{b}{\mu_p}+c,    
 \end{equation}
and an alternative hypothesis with an extra exponent parameter $a$:
 \begin{equation}
 \eta_1(\mu_p; a,b,c ) = \textstyle\frac{b}{\mu_p^a}+c.    
 \end{equation}
We will assume that ${\rm CV}^2_p$ is normally distributed about $\eta$ with unknown variance $\sigma^2_\eta$.

In this context, a maximum likelihood analysis is equivalent to least-squares analysis. Let the sum of the squares be defined:
\begin{equation}
 S_I({\bm \theta}) \equiv \sum_i [{\rm CV}_{p,i}^2- \eta_I(\mu_{p,i}; {\bm \theta} )]^2 
\end{equation}
for model $I$ where $\bm \theta$ represents the parameter vector. The maximum likelihood parameters are 
\begin{equation}
\hat{\bm \theta} = \arg \max_{\bm \theta} S_I({\bm \theta}),
\end{equation}
with residual norm:
\begin{equation}
\hat{S}_I = S_I(\hat{\bm \theta}).
\end{equation}
To test the null hypothesis, we will use the canonical likelihood ratio test with the test statistic:
\begin{equation}
\Lambda \equiv 2\log \frac{q_1}{q_0}, 
\end{equation}
where $q_0$ and $q_1$ are the likelihoods of the null and alternative hypotheses, respectively. Wilks' theorem states that $\Lambda$ has a chi-squared distribution of dimension equal to the difference of the dimension of the alternative and null hypotheses ($3-2=1$). 

\subsubsection{Hypothesis test I}

In our first analysis, we will estimate the variance directly. We computed the mean-squared difference for successive ${\rm CV}^2_p$ values, sorted by mean protein number $\mu_p$. The variance estimator is 
\begin{equation}
\hat{\sigma}^2_\eta = \textstyle\frac{1}{2} \left<({\rm CV}^2_{p,i}-{\rm CV}^2_{p,i+1})^2 \right>_i=6.3\times 10^{-4},
\end{equation}
where the brackets represent a standard empirical average over gene $i$ for the $\mu_p$-ordered gene ${\rm CV}_p^2$ values. The test statistic can now be expressed in terms of the residual norms:
\begin{eqnarray}
\Lambda &=& (\hat{S}_1-\hat{S}_2)/\hat{\sigma}^2_\eta, \\
&=& 3.3\times 10^4,
\end{eqnarray}
which corresponds to a p-value far below machine precision. We can therefore reject the null hypothesis.

\subsubsection{Hypothesis test II}

In a more conservative approach, we can use maximum likelihood estimation to estimate the variance of each model independently as a model parameter. In this case, the test statistic can again be expressed in terms of the residual norms:
\begin{eqnarray}
\Lambda &=& N \log \textstyle\frac{\hat{S}_1}{\hat{S}_2}, \\
&=& 1.6\times 10^2,
\end{eqnarray}
where $N$ is the number of data points. In this case, the p-value can be computed assuming the Wilks' theorem (\textit{i.e.}~the chi-squared test):
\begin{equation}
p = 6\times10^{-36},    
\end{equation}
again, strongly rejecting the null hypothesis.

\subsubsection{Maximum likelihood estimates of the parameters}

\label{sec:yeastnoiseempiricalmodel}

In the alternative hypothesis, the maximum likelihood estimate (MLE) of the empirical noise model (Eq. \ref{eqn:noiseempirical}) parameters are (Fig. \ref{fig:yeastnoise}):
\begin{eqnarray}
a &=& 0.57 \pm 0.02,\\
b &=& 3.0 \pm 0.5,\\        
c &=& 0.013  \pm 0.001,    
\end{eqnarray}
where the parameter uncertainty has been estimated using the Fisher Information in the usual way using the MLE estimate of the variance. 

\begin{figure}
    \centering
    \includegraphics[width=0.48\textwidth]{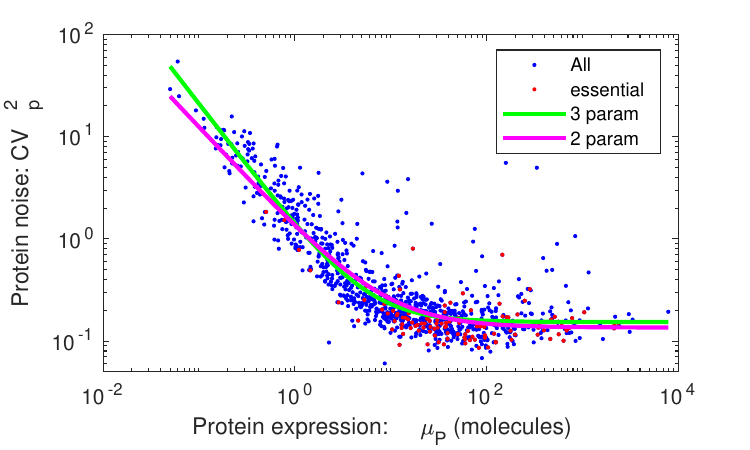}
    \caption{\textbf{Three-parameter fit to \textit{E.~coli}  noise. } The noise as a function of protein abundance from Taniguichi \textit{et al.} was fit to the 3 parameter noise model (Eqn. \ref{eqn:noiseempirical}). From the fit, protein noise scales proportionally with $\mu_p^{-1.22}$, which is a close result to the canonical model with $\mu_p^{-1}$. }
    
    \label{fig:EcoliNoise3param}
\end{figure}

The noise model parameters were also determined for \textit{E. coli}:
\begin{eqnarray}
a &=& 1.22 \pm 0.01,\\
b &=& 1.27 \pm 0.02,  \label{eqn:bparam}\\        
c &=& 0.154  \pm 0.002,    
\end{eqnarray}

with the corresponding fit shown in Fig. \ref{fig:EcoliNoise3param}. Since \textit{a} is close to 1, the canonical model with $a=1$ (Eqn. \ref{eqn:nullmodel}) is a somewhat reasonable approximation for the noise in \textit{E. coli}.

\subsubsection{Statistical details MLE estimate of the variance}

The minus-log-likelihood for the normal model $I$ is:
\begin{equation}
h_I(\hat{\bm \theta},\sigma^2) = \textstyle\frac{N}{2} \log 2\pi \sigma^2 + \textstyle\frac{1}{2\sigma^2}\hat{S}_I, 
\end{equation}
where $\hat{S}_I$ is the least-square residual. We then minimize $h_I$ with respect to the variance $\sigma^2$:
\begin{equation}
\partial_{\sigma^2} h|_{\hat{\sigma}^2} = 0,
\end{equation}
to solve for the MLE  $\hat{\sigma}^2$:
\begin{equation}
\hat{\sigma}^2 = \textstyle \frac{1}{N} \hat{S}_I.
\end{equation}
Next we evaluate $h$ at the variance estimator:
\begin{equation}
h_I(\hat{\bm \theta},\hat{\sigma}^2) = \textstyle\frac{N}{2} \left[\log 2\pi \frac{\hat{S}_I}{N} + 1\right].
\end{equation}
The test statistics can be written in terms of the $h$'s:
\begin{eqnarray}
  \Lambda &=& 2h_0(h_1(\hat{\bm \theta},\hat{\sigma}^2)-2h_2(\hat{\bm \theta},\hat{\sigma}^2),\\
  &=& N \log \textstyle \frac{\hat{S}_1}{\hat{S}_1},
\end{eqnarray}
which can be evaluated directly in terms of the residual norms for the null and alternative hypotheses.







\begin{table*}
\resizebox{\textwidth}{!}{\begin{tabularx}{1.15\textwidth}{ c | c | X  |c c c| }
Gene  & Message   & Annotated function & Essential (E)/   \\
name &  number:       &  from Ecocyc                  &   Nonessential (N)      \\ 
& $\mu_m$&  & Ref.~\cite{Baba:2008vn}, \cite{Gerdes:2003ys}, \cite{Goodall:2018rm} \\ \hline   
\textit{alsK} &        0.3  & The alsK gene encodes a D-allose kinase. Its role in the degradation of D-allose is unclear; AlsK is not required for utilization of a D-allose carbon source; this effect may be due to the presence of other ambiguous sugar kinases within \textit{E.~coli} K-12. &  E, N, N \\
& & & \\
    \textit{bcsB} &    0.4 & BcsB is encoded in a predicted operon together with \textit{bcsA}, \textit{bcsZ} and \textit{bcsC}. In other organisms, these genes are involved in cellulose biosynthesis, a characteristic of the rdar (red, dry and rough) morphotype. However, the K-12 laboratory strain of \textit{E.~coli} does not show a rdar morphotype and does not produce cellulose. &  E, N, N \\ 
    & & & \\
    \textit{entD} &    0.4 & AcpS is the founding member of a 4'-phosphopantetheinyl (P-pant) transferase protein family that includes \textit{E.~coli} EntD, \textit{E.~coli} o195 protein, and \textit{Bacillus subtilis} Sfp; family members share two conserved motifs but relatively low sequence identity overall. & E, N, N  \\
    & & & \\
    \textit{yafF} &    0.4 & No information about this protein was found by a literature search conducted on April 19, 2017. & E,-, N
 \\
    & & & \\
    \textit{yagG} &    0.6 & \textit{yagGH} is predicted to be a member of the XylR regulon; its products may mediate transport (YagG) and hydrolysis (YagH) of xylooligosaccharides; putative XylR and CRP binding sites are identified upstream of \textit{yagGH}. &  E,-, N \\
    & & & \\
    \textit{yceQ} &    0.2 & No information about this protein was found by a literature search conducted on July 12, 2017. & E, E, N \\
    & & & \\
    \textit{ydiL} &    0.2 & No information about this protein was found by a literature search conducted on April 7, 2017. & E, N, N \\ 
    & & & \\
    \textit{yhhQ} &   0.4 & YhhQ is an inner membrane protein implicated in the uptake of queuosine (Q) precursors - 7-cyano-7-deazaguanine (preQ0) and 7-aminomethyl-7-deazaguanine (\textit{preQ1}) - for Q salvage. Q-modified tRNA is absent in $\Delta$\textit{queD} and $\Delta$\textit{queD} $\Delta$\textit{yhhQ} strains grown in minimal media with glycerol; Q-modified tRNA is detected when a $\Delta$\textit{queD} strain is grown in minimal media plus 10 nM \textit{preQ0} or \textit{preQ1} but is absent when a $\Delta$\textit{queD} $\Delta$\textit{yhhQ} strain is grown under these conditions. \textit{yhhQ} expressed from a plasmid restores the presence of Q-modified tRNA in a $\Delta$\textit{queD} $\Delta$\textit{yhhQ} strain. &  E,-, N  \\
    & & & \\
    \textit{yibJ} &    0.3 & No information about this protein was found by a literature search conducted on July 9, 2018. & E, N, N \\
    & & & \\
    \textit{ymfK} &    0.4 & YmfK is a component of the relic lambdoid prophage e14 and is likely the SOS-sensitive repressor. It is similar to the P34 gene of the \textit{Shigella flexneri} bacteriophage SfV and belongs to the LexA group of SOS-response transcriptional repressors. & E, E, E 
 \end{tabularx}}

\caption{ \textbf{Below-threshold essential genes identified in \textit{E.~coli}.} This table describes the message numbers and annotations for essential genes that we estimated to have expression below the threshold of one message per cell cycle. However, in the final column, we show classifications from three different studies. Only one of the identified genes, \textit{ymfK}, was consistently defined as essential. } \label{ess_tab}

\end{table*}

\subsubsection{Detailed discussion of noise in \textit{E.~coli}}

\label{sec:noiseEcoli}

In general, the Telegraph model predicts that the noise will have a coefficient of variation \cite{Paulsson:2000xi,Friedman:2006oh}:
\begin{equation}
{\rm CV}_p^2 \approx \frac{1}{\mu_p} + \frac{\varepsilon \ln 2}{\mu_p}, \label{eqn:ecolinoise}
\end{equation}
where the first term is significant whenever the translation efficiency isn't $\varepsilon\gg 1$. In both \textit{E.~coli} ($\varepsilon \approx 30$) and yeast ($\varepsilon \approx 420$), this would seem naively to be the case. However, since translation efficiency in yeast is not uniform, we must consider its variation for low-expression proteins. We estimate that the detection efficiency in yeast is roughly $10^3$ molecules. Using Eq.~\ref{eqn:empmod}, we estimate that $\varepsilon\approx 100$ and our approximation holds  at the low-expression detection limit. 

In \textit{E.~coli}, the situation is somewhat more complicated. Unlike yeast, the translation efficiency is roughly constant (at high to intermediate expression levels) with respect to expression level \cite{Balakrishnan:2022ai}, and therefore both terms in Eq.~\ref{eqn:ecolinoise} are expected to scale like the canonical model ($\propto \mu_p^{-1}$). However, it is clear that the translation efficiency must significantly decrease for the lowest abundance proteins. This is visible even in  Ref.~\cite{Balakrishnan:2022ai} Fig.~1B, where the data falls below the predicted protein abundance at low message number. Note that these mass-spec measurements are not as sensitive as fluorescence-based measurements (\textit{e.g.}~only 64\% proteome could be detected \cite{Mori:2021ve}). Furthermore, fits to the \textit{E.~coli} noise (Eq.~\ref{eqn:bparam}) are consistent only with low values of $\varepsilon$.  
At sufficiently high expression levels such that we are confident about the translation efficiency, the noise is already very close to the noise floor.

\end{document}